\documentclass[12pt,a4paper]{article}
\usepackage{amsfonts,latexsym}
\usepackage{graphicx,color}
\usepackage{dcolumn}
\usepackage{graphicx}
\usepackage{amsmath}
\usepackage{amsfonts}
\usepackage{amssymb}
\usepackage{psfrag}
\usepackage{wrapfig}
\usepackage{subfigure}
\usepackage{makeidx}
\usepackage{caption}

\renewcommand{\thefootnote}{\fnsymbol{footnote}}

\begin{document}

\vspace{12mm}

\begin{flushright}
CTPU-PTC-25-34
\end{flushright}

\begin{center}
{{{\Large {\bf Negative potential-induced scalarization in  the Einstein-Euler-Heisenberg black hole}}}}\\[10mm]
{Hong Guo$^1$\footnote{e-mail address: guohong@ibs.re.kr}, Miok Park$^1$\footnote{e-mail address: miokpark76@ibs.re.kr}, and Yun Soo Myung$^{2}$\footnote{e-mail address: ysmyung@inje.ac.kr}}\\[8mm]

{${}^1$Particle Theory and Cosmology Group, Center for Theoretical Physics of the Universe, \\
 Institute for Basic Science (IBS), Daejeon 34126, Republic of Korea\\[0pt]}

{${}^2${Center for Quantum Spacetime, Sogang University, Seoul 04107, Republic of  Korea\\[0pt] }}

\end{center}
\vspace{2mm}

\begin{abstract}
We investigate a negative potential-induced scalarization of the Einstein-Euler-Heisenberg (EEH) black hole in the EEH-scalar (EEHS) theory, characterized by mass $M$, Euler–Heisenberg parameter $\mu$, and magnetic charge $q$. Within this framework, the charge $q$ can exceed the extremal bound \( q/M > 1 \), and a single event horizon is maintained provided the parameter \( \mu \) exceeds the \( \mu_{\text{max}} = 0.019 \), with the ADM mass fixed at $M=1/2$. We obtain a single branch of scalarized EEH (sEEH) black holes for $q>0$ which is considered as the simplest model for scalarization of EEH black holes.
We found that this class of hairy black holes is not thermodynamically favored, and their quasinormal modes indicate they are dynamically unstable. An interesting feature is that when $q<1/2$, the scalar charge varies only slightly with $q$ for a fixed mass. In contrast, for $q>1/2$, the scalar charge increases more rapidly as $q$ increases. This distinct behavior suggests that the scalar charge exhibits the characteristics of a primary charge for \( q < 1/2 \), and of a secondary charge for \( q > 1/2 \). This finding reveals notable features of hairy black holes in EEH theory, specifically in the overcharging regime. 
\end{abstract}

\vspace{1.5cm}

\hspace{11.5cm}
\newpage
\renewcommand{\thefootnote}{\arabic{footnote}}
\setcounter{footnote}{0}

\section{Introduction}

A fundamental question in black hole physics is whether these objects can possess rich structures, often referred to as ``hair." Following the development of the Uniqueness Theorem \cite{ Israel:1967wq, Israel:1967za}, the no-hair conjecture was proposed, stating that a stationary, axisymmetric black hole is fully described by its mass, electric charge, and angular momentum \cite{Ruffini:1971bza}. This conjecture was later proven more rigorously by J. Bekenstein for the specific case of scalar fields, demonstrating that black holes cannot sustain scalar ``hair" under certain conditions \cite{Bekenstein:1972ny, Bekenstein:1995un}. However, evasions of the no-hair theorem have been found in theories of gravity that extend beyond general relativity (GR) or incorporate nonlinear matter fields. For a comprehensive review of these cases, we refer the reader to Refs.~\cite{Herdeiro:2015waa,Volkov:2016ehx}.

 Under these circumstances, various black holes with scalar hair have been numerically constructed in Einstein-Gauss-Bonnet-scalar theories~\cite{Doneva:2017bvd,Silva:2017uqg,Antoniou:2017acq,Latosh:2023cxm} and Einstein-Maxwell-scalar (EMS) theories~\cite{Herdeiro:2018wub}. These solutions have inspired further studies aiming to understand the formation mechanisms of scalar hair from initially bald black holes. One such mechanism, known as spontaneous scalarization \cite{Silva:2017uqg} suggests that a tachyonic instability—induced by an effective negative mass squared of the scalar perturbation—triggers the growth of a dynamical scalar field from a constant scalar background. However, the stability of the resulting hairy black holes must be examined independently each case. On the other hand, another work \cite{Latosh:2023cxm,Hyun:2024sfv} employs a scalar coupling function with mass and quartic interaction terms, demonstrating the possibility of a phase transition from black holes to hairy black holes via spontaneous symmetry breaking in a symmetry-broken phase. This alternative mechanism has the advantage of ensuring the stability of the final state.

An intriguing feature of the EMS theory \cite{Herdeiro:2018wub} is the existence of overcharged black holes with charge-to-mass ratios $q\le M$ which is significantly distinct from Reissner-Nordstr\"{o}m (RN) black holes. The similar feature appears in gravitational theories coupled to nonlinear electrodynamics.
Euler and Heisenberg (EH) proposed a novel framework in which one-loop corrections are incorporated into quantum electrodynamics (QED), introducing a EH parameter $a=8\alpha_{fs}/45m^4(\equiv8\mu)$ to explain the vacuum polarization in QED and control the strength of the nonlinear electrodynamics contribution~\cite{Heisenberg:1936nmg}. 
Yajima and Tamaki later derived the Einstein-Euler-Heisenberg (EEH) black hole solution by considering  the one-loop effective Lagrangian together with the Einstein gravity~\cite{Yajima:2000kw}. 
In this theory, the black hole charge can naturally extend into the regime $q \geq M$ without any restrictive assumptions.
It should be emphasized that the physics of black holes in the overcharged regime remains largely unexplored. 
This is an important open question, as it touches on issues related to naked singularities and the weak cosmic censorship conjecture~\cite{Hubeny:1998ga,Liu:2020cji,Alipour:2025dan,Kehagias:2023qmy,Shaymatov:2023gfh}. 
In the context of the EEH gravity framework, it is also closely connected to the potential detection of quantum gravity effects~\cite{Brodin:2001zz,Allahyari:2019jqz,Kruglov:2020tes}.


In the EEH framework, it is interesting to explore whether black holes can support scalar hair. Analytical solutions for charged EEH black holes with scalar hair were recently found by introducing a specific scalar potential, with a scalar field profile characterized by $\phi(r) \sim \ln(1+\nu/r)$ \cite{Karakasis:2022xzm}. In this present work, we aim to study the scalarization of magnetically charged EEH black holes by employing a negative scalar potential, $V(\phi) = -\alpha^2 \phi^6$. This potential is particularly useful because it is known to be one of the simplest forms that can give rise to scalarized black holes, even in the Einstein–minimally coupled scalar theory \cite{Chew:2024evh}. We try to obtain scalarized EEH black holes (sEEH) and then analyze the distinct properties exhibited by magnetic charge within different parameter regimes.


Our results reveal that a negative scalar potential can indeed trigger the formation of sEEH black holes, providing yet another example of a no-hair theorem evasion~\cite{Herdeiro:2015waa}. By solving the equations of motion, we obtain such a single branch of sEEH black hole solutions. We found that when the magnetic charge \( q < 1/2 \), the scalar charge varies only slightly with \( q \) for fixed \( M \), whereas for \( q > 1/2 \), it increases more rapidly as \( q \) increases. This distinct behavior suggests that the scalar charge exhibits the characteristics of a primary charge for \( q < 1/2 \), and of a secondary charge for \( q > 1/2 \). It would be interesting to examine more precisely in future whether the scalar charge is truly primary or not. We also investigated thermodynamic properties and the stability of these hairy black holes by analyzing their scalar perturbations and quasinormal modes (QNMs). Our findings indicate that this class of hairy black holes is neither thermodynamically favored nor dynamically stable. Nevertheless, these solutions provide a valuable theoretical framework for exploring nontrivial scalar field configurations on black holes. The negative potential itself might offer a mechanism for stable hairy black holes to lose their scalar hair. 

\section{EEHS theory and its black hole stability}

We introduce  the Einstein-Euler-Heisenberg-scalar (EEHS) theory with  a term $\mathcal{F}^2$
\begin{equation}
S_{\rm EEHS}=\frac{1}{16 \pi}\int d^4 x\sqrt{-g}\Big[ R-2\partial_\mu \phi \partial^\mu \phi-V(\phi)- \mathcal{F}+\mu \mathcal{F}^2\Big],\label{Act1}
\end{equation}
where we notify the Maxwell term  $\mathcal{F}=F_{\mu\nu} F^{\mu\nu}$, the scalar potential $V(\phi)=-4\alpha^2 \phi^6$, and the EH  parameter $\mu$.
The bald black hole solutions including $\mu$ were discussed extensively in the  EEH theory without scalar counterpart~\cite{Yajima:2000kw,Allahyari:2019jqz,Amaro:2020xro,Breton:2021mju}.

We  derive  the Einstein  equation from the action \eqref{Act1}
\begin{eqnarray}
 G_{\mu\nu}\equiv2(T^{\phi}_{\mu\nu}+T^{F}_{\mu\nu}), \label{equa1}
\end{eqnarray}
where two energy-momentum tensors take the forms
\begin{eqnarray}
T^\phi_{\mu\nu}&=&\partial _\mu \phi\partial _\nu \phi -\frac{1}{2}\Big[(\partial \phi)^2+\frac{V(\phi)}{2}\Big]g_{\mu\nu}, \label{emten0} \\
T^F_{\mu\nu}&=&F_{\mu\rho}F_{\nu}~^\rho-\frac{1}{4}\mathcal{F}g_{\mu\nu}
          -2\mu \mathcal{F}F_{\mu\rho}F^\rho_\nu+\frac{\mu}{4}\mathcal{F}^2g_{\mu\nu} \label{emten}.
\end{eqnarray}
The Maxwell equation leads to
\begin{eqnarray} \label{M-eq}
&&\nabla_\mu (F^{\mu\nu}-2\mu\mathcal{F}F^{\mu\nu})=0.\label{M-eq1}
\end{eqnarray}
Interestingly, the scalar equation takes a simple form, characteristic of a minimally coupled scalar theory
\begin{equation}
\square \phi=\frac{V'(\phi)}{4} \label{s-equa}.
\end{equation}
First of all, let us introduce  the mass function $\bar{m}(r)$ together with the magnetic field strength $\bar{\mathcal{F}}=2q^2/r^4$ and  $\bar{\phi}=0$. In this case, the ${T}_t~^t$-component of Einstein equation reduces to
\begin{equation}
\bar{m}'(r)=\frac{q^2}{2r^2}-\mu \frac{q^4}{r^6}. \label{mass-eq}
\end{equation}
Solving this leads to the EEH black hole solution
\begin{eqnarray}
ds^2_{\rm EEH}=\bar{g}_{\mu\nu}dx^\mu dx^\nu=-f(r) dt^2+\frac{dr^2}{f(r)} +r^2d\Omega^2_2   \label{EEH-s}
\end{eqnarray}
with
\begin{equation} \label{metric-func}
 f(r)\equiv1-\frac{2\bar{m}(r)}{r}=1-\frac{2M}{r}+\frac{q^2}{r^2}-\frac{2\mu q^4}{5r^6}.
\end{equation}
This black hole solution could be described by a set of three parameters $(M,q,\mu)$, where $M$ denotes the ADM mass, $q$ is the magnetic charge, and $\mu$ is the EH parameter.

\begin{figure*}[t!]
   \centering
  \includegraphics[width=0.4\textwidth]{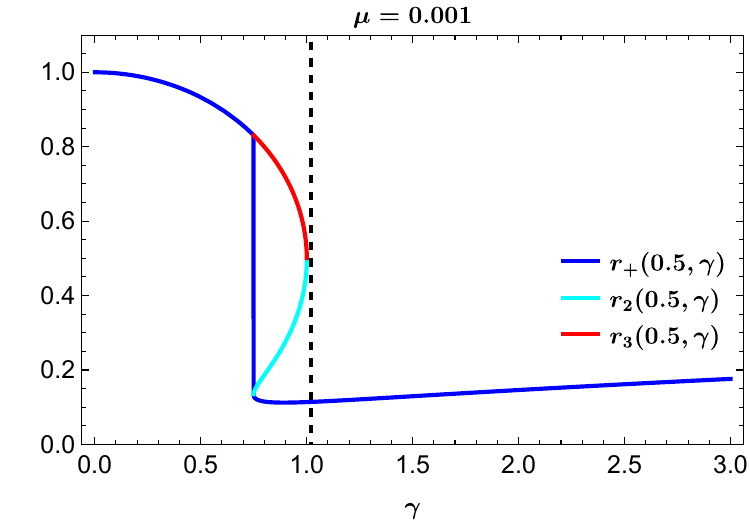} \qquad \qquad \qquad
  \includegraphics[width=0.4\textwidth]{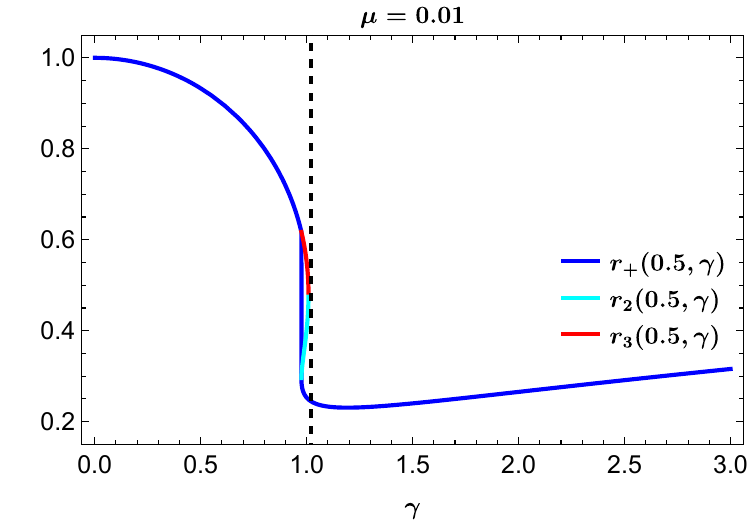}
  \includegraphics[width=0.4\textwidth]{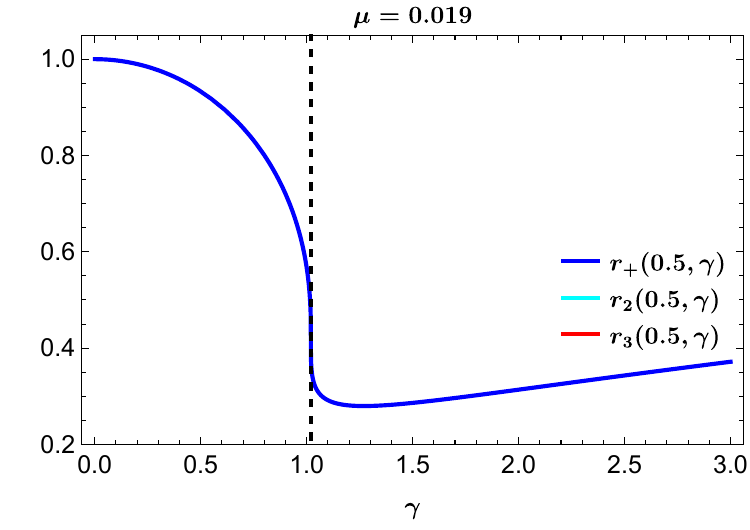} \qquad \qquad \qquad
  \includegraphics[width=0.4\textwidth]{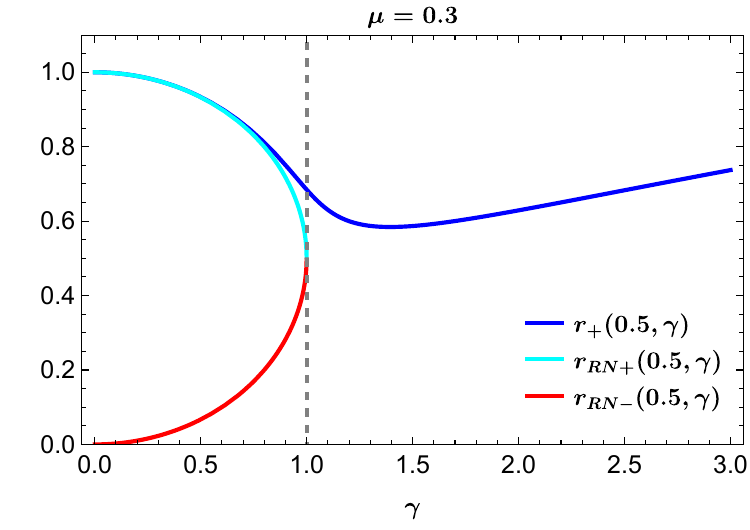}
\caption{The parameter $\gamma$ represents the charge-to-mass ratio, with the dashed black line indicating $\gamma=1.02$. In the EEH black hole, three distinct real positive horizons exist only when $\mu \lesssim 0.019$, as shown in the top and bottom-left figures. When a single horizon exists $\mu > 0.019$, its size is compared with that of a RN black hole of the same ADM mass ($M=0.5$) in the bottom-right figure.
\label{fig:EEHhorizon}
}
\end{figure*}

Assuming $q>0$, $M>0$, and $\mu>0 $, the solution yields three real positive roots under the following conditions:
\begin{itemize}
    \item[a.] $(0< \gamma <1) \; \; $  $0 < \mu <x_1 + x_2$, $\qquad$ or
    \item[b.] $(1 < \gamma <1.02) \; \; $  $x_1-x_2 < \mu < x_1+x_2$
\end{itemize}
where we introduce the charge-to-mass ratio by $q = \gamma M$, while $x_1$, $x_2$ are given by
\begin{align} 
x_1 =& \bigg(-\frac{5.35837}{\gamma ^4}+0.37037 \gamma ^2+\frac{9.64506}{\gamma ^2}-4.62963 \bigg) M^2, \\
x_2=& \frac{0.00171468 M^2}{\gamma^4}\bigg(-1.11974\times 10^6 \gamma ^{10}+9.72\times 10^6 \gamma ^8-3.1725\times 10^7 \gamma ^6 \nonumber\\
&+4.85156\times 10^7 \gamma ^4-3.51563\times 10^7 \gamma ^2+9.76563\times 10^6 \bigg)^{\frac{1}{2}}.
\end{align}
If $M=0.5$ is fixed, the square root in \( x_2 \) requires \( 0 < \gamma \lesssim 1.02 \). Since \( x_1 + x_2 \) increases with increasing \( \gamma \), the parameter \( \mu \) reaches its maximum value, \( \mu_{\textrm{max}} \sim 0.019 \), at \( \gamma \sim 1.02 \), corresponding to the point where two roots become degenerate. Therefore, three real positive roots exist only when the charge-to-mass ratio satisfies \( 0 < \gamma \lesssim 1.02 \). If \( \mu \) exceeds \( \mu_{\textrm{max}} \sim 0.019 \), then only a single real positive root exists for all values of \( \gamma \). This corresponds to the case where the charge-to-mass ratio is unbounded, in sharp contrast to the RN black hole (\( \mu = 0 \), i.e., without QED effects), which possesses outer and inner horizons given by \( r_{\mathrm{RN}\pm}(M, q) = M \pm \sqrt{M^2 - q^2} \). As demonstrated in Fig.~\ref{fig:EEHhorizon}, we set $\mu=0.001, 0.01, 0.19$, and $0.3$ as examples to illustrate this relationship.


Hereafter, we fix the EH parameter $\mu= 0.3$ so as to obtain a black hole with a single horizon.
In this setting, we do not worry about the no scalar-haired inner horizon theorem, which corresponds to no-go theorem for the scalar-haired Cauchy horizon~\cite{Devecioglu:2021xug}.
This is one of the motivations for studying black hole scalarization in the regime of unbounded magnetic charge.


Before proceeding, we derive the temperature from the surface gravity as
\begin{equation}
    \frac{f'(r)}{4\pi}|_{r\to r_+} \to \tilde{T}(M,q)=\frac{1}{4\pi}\Big[\frac{0.6 q^4}{r_+^7(M,q)}-\frac{q^2}{r_+^3(M,q)}+\frac{1}{r_+(M,q)}\Big],
\end{equation}
where the last two terms represent the temperature $\tilde{T}_{\rm RN}(M,q)$ for the RN black hole (see Fig.~5 (Left)). 
Also, the area-law entropy for EEH black hole is defined by 
\begin{equation}
    \tilde{a}_H=\pi r_+^2(M,q)
\end{equation}
with its minimum of 1.06 occurring at $q=0.7$ with $M=0.5$ (see Fig.~5 (Right)).

We now introduce perturbations around the EEH black hole
\begin{equation}
g_{\mu\nu}=\bar{g}_{\mu\nu}+h_{\mu\nu},\quad\phi=0+\delta \varphi, \quad F_{\mu\nu}=\bar{F}_{\mu\nu}+f_{\mu\nu}
\end{equation}
with a metric perturbation $h_{\mu\nu}$ and a linearized Maxwell field $f_{\mu\nu} =\partial_\mu a_\nu-\partial_\nu a_\mu$.
The linearized scalar equation takes the form
\begin{equation}
\bar{\square}\delta \varphi=0.\label{per-eq}
\end{equation}
Our task is to solve the linearized scalar equation \eqref{per-eq} to examine the (in)stability of the EEH black hole.
This is because its linearized metric theory has been shown to be stable against metric perturbations around the EEH black holes~\cite{Luo:2022gdz}.
The latter conclusion was reached through the computation of QNM frequencies for both axial and polar metric perturbations, which differs from those of RN black holes.

\begin{figure*}[t!]
   \centering
  \includegraphics[width=0.4\textwidth]{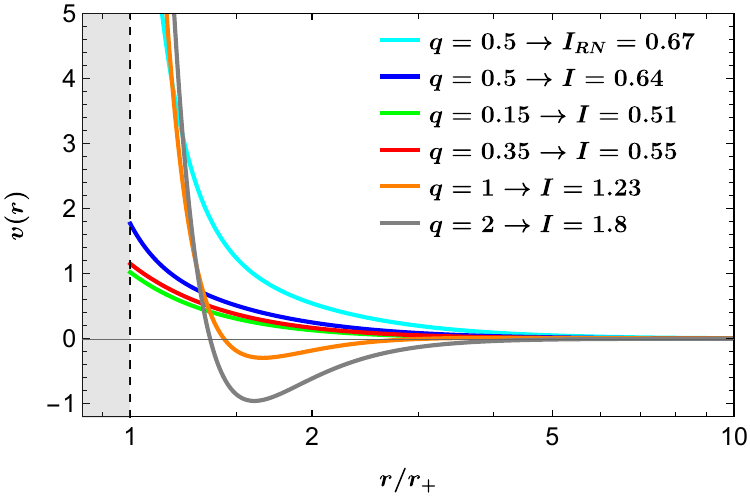}
   \hfill%
\includegraphics[width=0.4\textwidth]{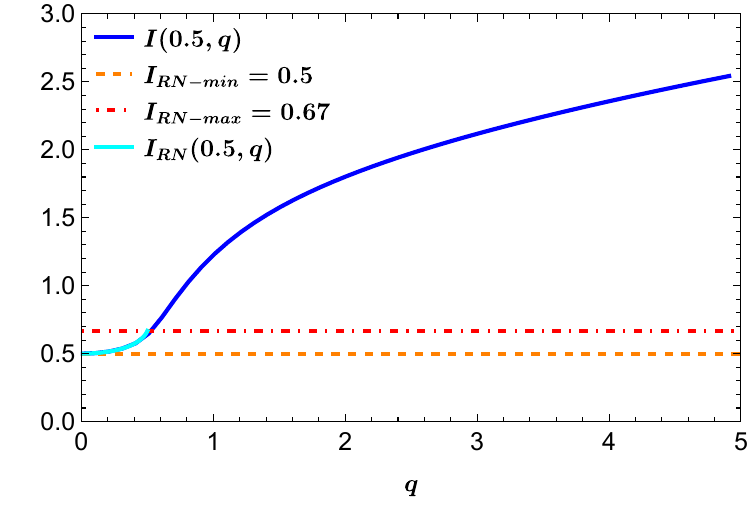}
\caption{Potential $v(r,M,q)$ and its integration $I(M,q)$.
(Left) Five $q$-dependent potentials $v(r,M=0.5,q)$ as functions of
$r/r_{+}$ with $q=0.15,0.35,0.5,1,2$. 
Negative regions appear near the horizon for $q>0.5$.
However, $v_{\rm RN}(r,0.5,q)$ is a positive function of $r\in[r_{\rm RN+},\infty]$ for $q\in[0,0.5]$.
(Right) Integration $I(0.5,q)$ as a function of $q$. 
Its lower limit is 0.5 at $q=0.0001$ and then, $I$ grows monotonically with increasing $q$. 
We note that $I_{\rm RN}(0.5,0)=0.5$ is considered as the minimum and $I_{\rm RN}(0.5,0.5)=0.67$ is regarded as its maximum.}
\end{figure*}

Considering a separation of variables
 \begin{equation}
 \delta \varphi(t,r,\theta,\hat{\phi}) =\int \sum_{lm} \varphi(r) Y_{lm}(\theta) e^{i m\hat{ \phi}} e^{-i\omega t} d\omega,\quad \varphi(r)=\frac{u(r)}{r},
\end{equation}
which transforms Eq. \eqref{per-eq} into a Schr\"odinger-type equation in terms of the tortoise coordinate $r_*=\int \frac{dr}{f(r)}$
\begin{equation}
\frac{d^2 u(r)}{dr^2_*} +\Big[\omega^2-V_{\rm EEH}(r,M,q)\Big] u(r) =0. \label{sch-eq}
\end{equation}
Here, the effective scalar potential is given by
\begin{eqnarray}
V_{\rm EEH}(r,M,q)&=&f(r)\Big[\frac{l(l+1)}{r^2}+\frac{2M}{r^3}-\frac{2q^2}{r^4}+\frac{3.6 q^4}{5r^8}\Big].\label{EEH-P}
\end{eqnarray}
In what follows, we focus on the $s(l=0)$-mode scalar perturbation for further analysis (see Fig.~2 (Left) for its $q$-dependent potential profiles $v(r,M,q)$).
A sufficient condition for instability is given by~\cite{Dotti:2004sh}
\begin{equation}
\int_{r_+(M,q)}^\infty \Big[\frac{V_{\rm EEH}(r,M,q)}{f(r)}\Big]dr=\int_{r_+(M,q)}^\infty v(r,M,q)dr\equiv I(M,q) <0.
\end{equation}
Although $v(r,0.5,q)$ develops negative regions near the horizon as $q$ increases, these negative contributions are insufficient to trigger an instability. 
Indeed, Fig.~2 (Right) shows clearly that $I(0.5,q)\ge 0.5$ for all $q>0$, confirming the stability of the EEH black hole. 
For RN black hole with $\mu=0$, $v_{\rm RN}(r,0.5,q)$ remains strictly positive for $r\in[r_{\rm RN+},\infty]$ within $q\in[0,0.5]$. 
Its integral attains a minimum value of 0.5 at $q=0$ and a maximum of 0.67 at $q=0.5$. 

\section{sEEH black holes and their property}

Let us derive sEEH black holes through negative potential-induced scalarization by solving the full set of equations.
For this purpose, we adopt the following metric and field ansatz~\cite{Herdeiro:2018wub}
\begin{eqnarray}\label{nansatz}
ds^2_{\rm sEEH}&=&-N(r)e^{-2\delta(r)}dt^2+\frac{dr^2}{N(r)}+r^2(d\theta^2+\sin^2\theta d\hat{\varphi}^2) \nonumber \\
N(r)&=&1-\frac{2m(r)}{r},\quad \phi=\phi(r),\quad A= A_{\hat{\varphi}} d\hat{\varphi}.
\end{eqnarray}
Substituting the gauge field ansatz into Eq.~\eqref{M-eq} yields the magnetic  potential $A_{\hat{\varphi}}=-q \cos \theta$, which in turn gives $F_{\theta \hat{\varphi}}=q\sin\theta$ and $\mathcal{F}=2q^2/r^4$.
This implies that no approximation for $A_{\hat{\varphi}}$ is required, the nonlinear Maxwell equation is solved exactly in this case.
Plugging Eq.\eqref{nansatz} into Eqs.\eqref{equa1} and \eqref{s-equa}, one finds three equations for $m(r),~\delta(r),$ and $\phi(r)$ as
\begin{eqnarray}
&&m'(r)=\frac{q^2}{2r^2}-\frac{0.3 q^4}{r^6}+\frac{r^2}{2}\Big[\Big(1-\frac{2m(r)}{r}\Big)\phi'^2(r)-\frac{V(\phi)}{2}\Big], \label{neom1}\\
&&\delta'(r)=-r\phi'^2(r), \label{neom2}\\
&&\frac{2}{r^2}\Big[m(r)+rm'(r)-r\Big]\phi'(r)+\Big(1-\frac{2m(r)}{r}\Big)[\delta'(r)\phi'(r)-\phi''(r)]=-\frac{V'(\phi)}{4}, \label{neom3}
\end{eqnarray}
where the prime ($'$) denotes differentiation with respect to its argument. 
We note that Eq.~\eqref{neom1} reduces to Eq.~\eqref{mass-eq} when $\phi(r)=0$.
Considering the existence of a single horizon located at $r=r_+$, the near-horizon behavior of the solution to Eqs.~\eqref{neom1}-\eqref{neom3} can be expressed as
\begin{eqnarray}
m(r)&=&\frac{r_+}{2}+m_1(r-r_+)+\cdots,\quad
\delta(r)=\delta_1(r-r_+)+\cdots,\label{aps-1}\\
\phi(r)&=&\phi_0+\phi_1(r-r_+)+\cdots,\label{aps-2}
\end{eqnarray}
where the expansion coefficients are given by
\begin{eqnarray}
&&m_1=\frac{q^2}{2r_+^2}-\frac{0.3 q^4}{r_+^6}+\frac{r_+^2V(\phi_+)}{4},\quad \delta_1=-r_+\phi_1^2,\quad \phi_1=\frac{r_+V'(\phi_+)}{4(1-2m_1)}.
\end{eqnarray}
Here the constant scalar $\phi_0\equiv \phi(r_+)$ will be
determined when matching with an asymptotically flat solution in the far region
\begin{eqnarray}
&&m(r)=M-\frac{q^2+q_s^2}{2r}+\cdots,\quad
\delta(r)=\delta_0+\frac{q_s^2}{2r^2}+\cdots,\quad
\phi(r)=\frac{q_s}{r}+\cdots,
\end{eqnarray}
where $M$ is the ADM mass, $q$ is the magnetic charge, and $q_s$ denotes the scalar charge. 
\begin{figure*}[t!]
   \centering
  \includegraphics[width=0.4\textwidth]{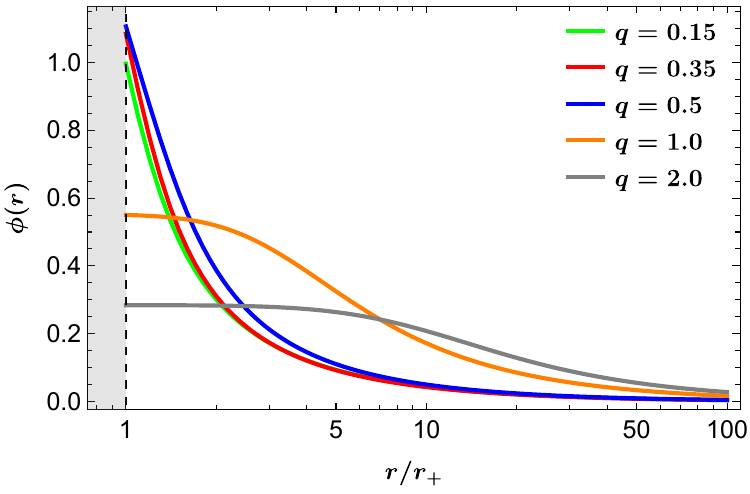}
 \hfill%
  \includegraphics[width=0.4\textwidth]{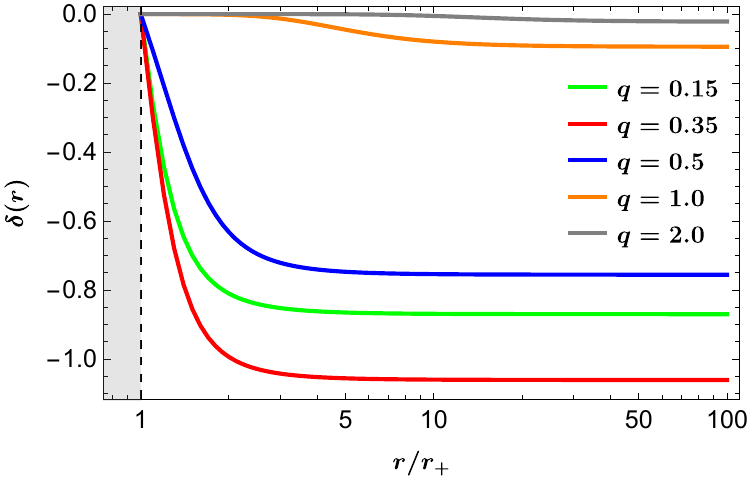}
\caption{(Left) Scalar hair $\phi(r\in[r_+,100],q)$ with $q=0.15,0.35,0.5,1,2$. These all are decreasing functions of $r$ with different initial values at $r=r_+$. (Right) $\delta(r\in[r_+,100],q)$ with different $q=0.15,0.35,0.5,1,2,$ are negative decreasing functions of $r$.}
\end{figure*}

Regarding explicit sEEH black hole solutions with $q=0.15,0.35,0.5,1,2$, we present numerical results for the single branch with $M=0.5,\mu=0.3,\alpha=1$.
As shown in Fig.~3, the scalar field $\phi(r,q)$ decreases monotonically with $r$, while $\delta(r,q)$ are negative monotonically decreasing functions of $r$.
Near the horizon, $\phi(r,q)$ is determined by three free parameters: $r_+$, $q$, and $\phi_0$. 
Using the shooting method to enforce $\phi(r \to \infty,q) = 0$ and $m(r \to \infty,q) = 1/2$, we can fix both $\phi_0$ and $r_+$.  
As a result, all numerical solutions are effectively characterized by tuning a single free parameter $q$. 
On the other hand, by setting $\delta(r_+,q) = 0$, $\delta(r,q)$ decreases monotonically in the near-horizon region. 
At large $r$, the asymptotic behavior of $\delta(r,q)$ depends on whether $q \leq 0.5$ or $q > 0.5$, consistent with the behavior of $\phi(r,q)$. 
These functions vanish identically for EEH black holes. 

For comparison, Fig.~4 display the two mass functions $\tilde{m}(r,q)$ for EEH black hole and $m(r)$ for sEEH black hole with $q=0.15,0.35,0.5,1,2$. 
In both cases, the mass functions asymptotically approach $M=0.5$ at large $r$, and a negative region emerges for $q\gtrsim 1$.
Although the sEEH black holes exhibit a deeper potential well compared to the EEH black holes with the same $q$, the two solutions remain qualitatively similar.
 \begin{figure*}[t!]
   \centering
  \includegraphics[width=0.4\textwidth]{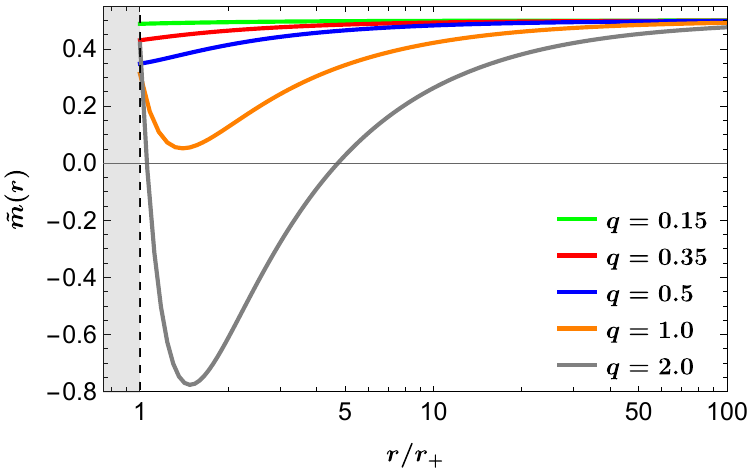}
 \hfill%
  \includegraphics[width=0.4\textwidth]{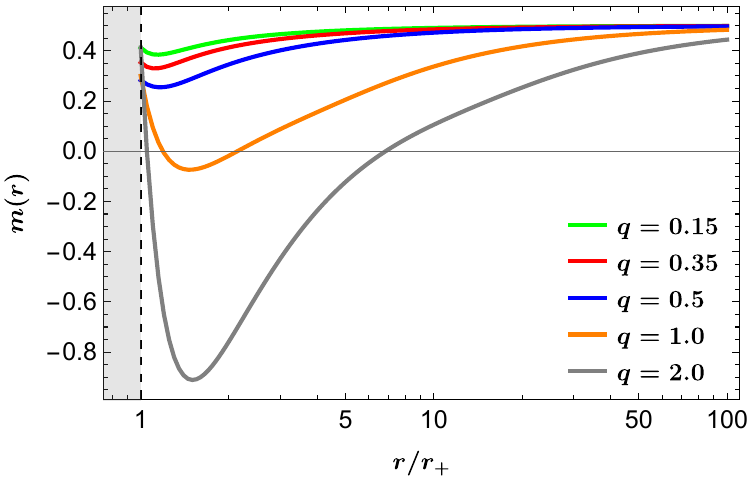}
\caption{(Left) Mass function $\tilde{m}(r\in[r_+,100],M=0.5,q)$ with $q=0.15,0.35,0.5,1,2$ for EEH black hole. They all converge $M=0.5$ at large $r$. Negative regions appear for $q=2$. (Right) Mass function $m(r\in[r_+,100],q)$ for sEEH black hole with $\alpha=1$. Negative regions appear for $q=1,2$. }
\end{figure*}
\begin{figure*}[t!]
   \centering
  \includegraphics[width=0.4\textwidth]{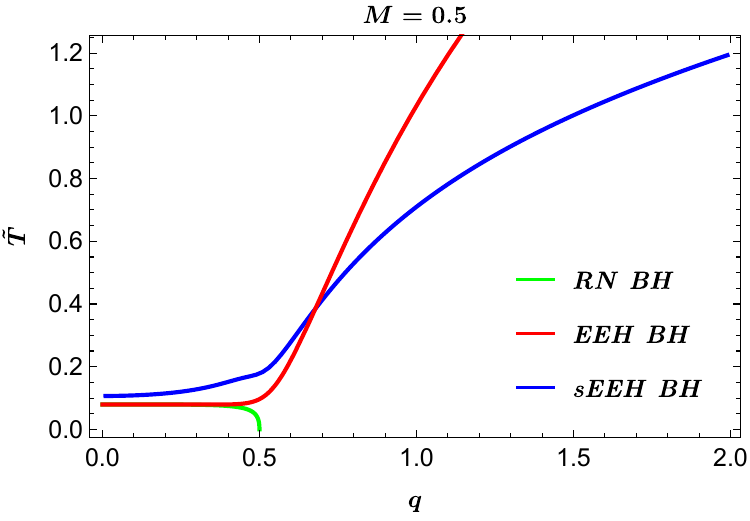}
 \hfill%
  \includegraphics[width=0.4\textwidth]{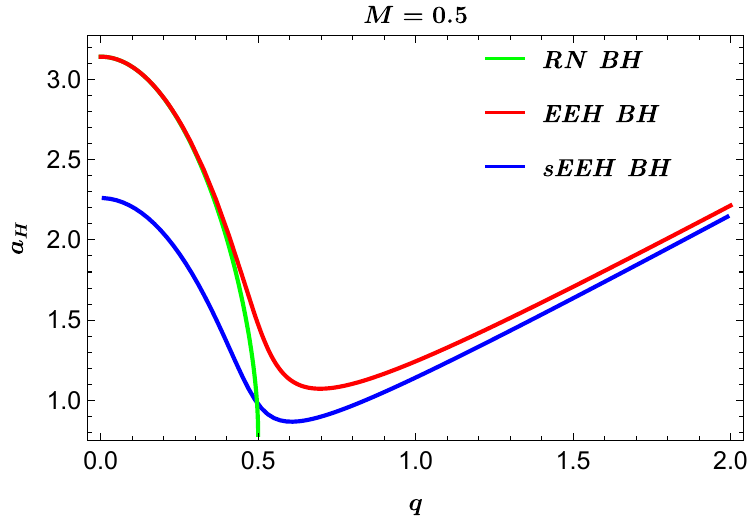}
\caption{(Left) Temperature $\tilde{T}(M=0.5,q)$, $T(M=0.5,q)$, and $\tilde{T}_{\rm RN}(M=0.5,q)$ as functions of $q$ for EEH, sEEH and RN ($\mu=0$) black holes. The first two are increasing functions while the last is a decreasing function.
The minimum temperature (=0.08) for EEH black hole is located at $q=0.35$, while the minimum temperature (=0) for RN black hole is at extremal point ($q=0.5$).  (Right) Area-law entropy  of the outer horizon as functions of $q$ with $M=0.5$  for EEH, sEEH and RN ($\mu=0$) black holes. The minimum entropy (=1.06) is located at $q=0.7$ for EEH black holes.  }
\end{figure*}

In Fig.~5 (Left), we present three temperature profiles for EEH, sEEH, and RN black holes as functions of $q$.
The temperatures $\tilde{T}(M=0.5,q)$ and $\tilde{T}_{\rm RN}(M=0.5,q)$ correspond to the EEH and RN black holes, respectively, with the former increasing and latter decreasing as functions of $q$. 
The EEH black hole reaches its minimum temperature $\tilde{T}=0.08$ at $q=0.35$, whereas the RN black hole's minimum temperature $\tilde{T}_{\rm RN}$=0 at extremal point ($q=0.5$).
The temperature for sEEH black holes is defined by
\begin{equation}
T(M,q)= \frac{1}{4\pi}N'(r_+)e^{-\delta(r_+)},   \label{s-temp}
\end{equation}
which increases with $q$, resembling the behavior of $\tilde{T}(M,q)$ for EEH black hole.
Its minimum temperature occurs near $q\simeq 0$. 

Fig.~5 (Right) shows the area-law entropy of the outer horizon as a function of $q$ with $M=0.5$.
It is defined as
\begin{equation}
    a_H(M,q)=\pi r_+^2. \label{s-ent}
\end{equation}
Here we observe the inequality $a_H(M=0.5,q)<\tilde{a}_H(M=0.5,q)$ for $q>0$, implying that the sEEH black holes represent less thermodynamically favorable configurations compared to the EEH black holes. 
Within the restricted interval $0<q<0.5$, the entropy ordering is $a_H< \tilde{a}^{\rm RN}_H < \tilde{a}_H$, which suggests that the EEH black hole remains the most thermodynamically preferred state in this range. 
The minimum entropy for EEH black hols is 1.06 at $q=0.7(>0.5)$, whereas the sEEH black hole attain a lower minimum entropy of 0.87 located at $q=0.61(<0.7)$. 
For RN black holes, the minimum entropy 0.79 occurs at $q=0.5$, while its maximum 3.14 is at $q=0$.

 \begin{figure*}[t!]
   \centering
  \includegraphics[width=0.4\textwidth]{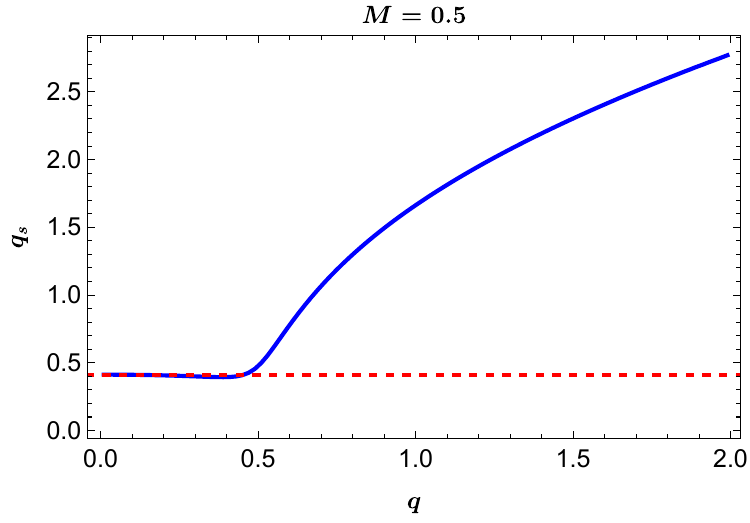}
 \hfill%
  \includegraphics[width=0.4\textwidth]{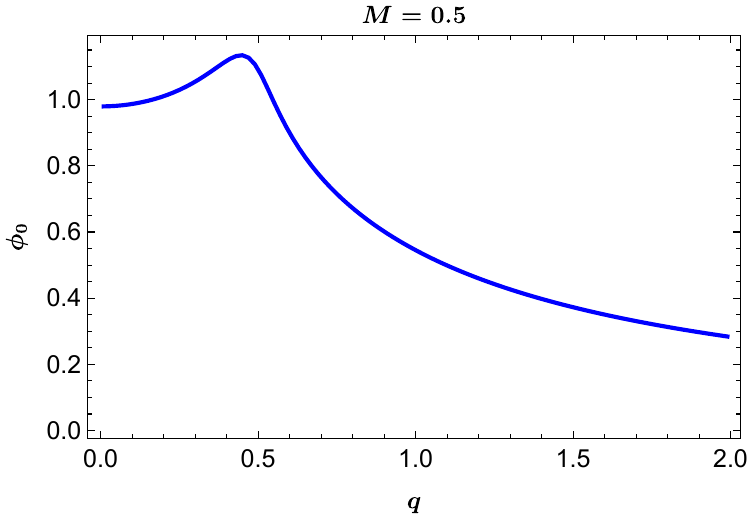}
\caption{(Left) Constant scalar charge and scalar charge $q_s(M=0.5,q)$ as an increasing function of $q$ for sEEH black holes.
(Right) Scalar constant $\phi_0(M=0.5,q)$ starting with an increasing function of $q$ and then, it decreases for larger values of $q$ in sEEH black holes.}
\end{figure*}

In Fig.~6, we display the scalar charge $q_s(M=0.5,q)$ and scalar constant $\phi_0(M=0.5,q)$ as functions of $q$. 
We found that $q_s(M=0.5,q)$ varies only slightly with $q$ for $q<0.5$, while it increases more rapidly when $q>0.5$.
This behavior corresponding to the differing scalar slops observed for various $q$ in Fig.~3 (Left).
Notably, this suggests that the scalar field exhibits the characteristics of a primary charge for $q \leq 0.5$, whereas the growth of $q_s(q)$ for $q > 0.5$ indicates a secondary scalar charge.
This result represents a novel feature of sEEH black holes.

On the other hand, the scalar constant $\phi_0(M=0.5,q)$ increases with $q$ for $q<0.5$, attains a maximum near $q=0.5$, and then decreases for $q>0.5$. 
This behavior reflects the distinct initial scalar hair configurations at the horizon $r/r_+=1$, as shown in Fig.~3 (Left).

\begin{figure*}[t!]
   \centering
  \includegraphics[width=0.4\textwidth]{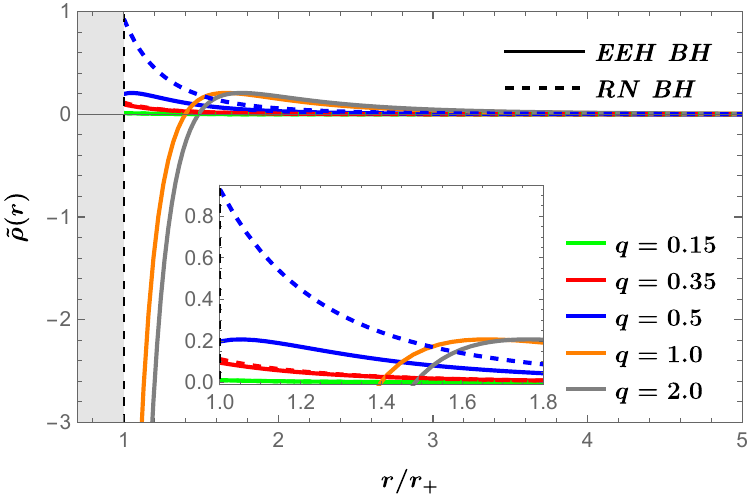}
   \hfill%
\includegraphics[width=0.4\textwidth]{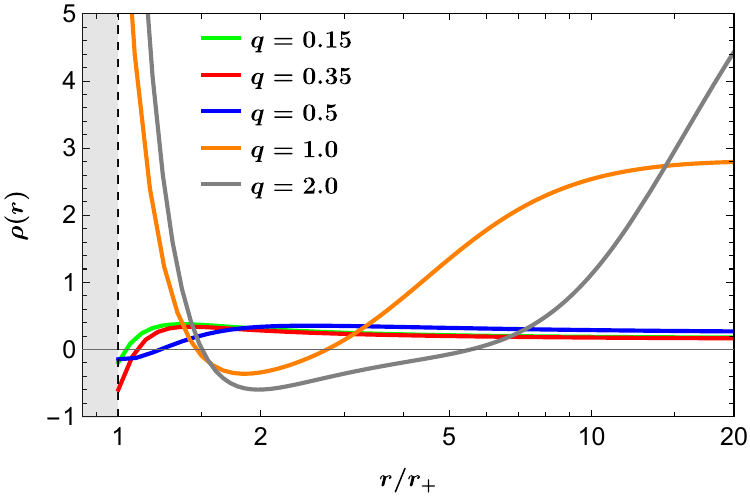}
\caption{Energy density $\rho(r,q)$ as a function of $r/r_+$.
(Left) Five profiles $\tilde{\rho}(r,q)$ as functions of
$r/r_+\in[1,5]$ with $q=0.15,0.35,0.5,1,2$ for EEH black holes, together with the energy density of RN black hole ($\mu=0$) $\tilde{\rho}_{\rm RN}(r,q)=q^2/2r^4$ over $r/r_+\in[1,5]$. 
Negative regions appear near the horizon for $q>0.5$.
In contrast, $\tilde{\rho}_{\rm RN}(r,0.5)$ remains positive for all $r$ and is nearly same with $\tilde{\rho}(r,0.5)$ for $r/r_+>2$.
(Right) Five profiles $\rho(r,M=0.5,q)$ as functions of
$r/r_+\in[1,20]$ with $q=0.15,0.35,0.5,1,2$ for sEEH black holes. 
All cases exhibit negative energy density regions.}
\end{figure*}

Finally, we consider the energy density $\rho(r)$ to preliminarily assess the stability of sEEH black holes
\begin{equation} \label{rho-1}
    \rho(r)=-T^{\phi~t}_t-T^{F~t}_t=\frac{N(r)\phi'^2(r)}{2}-\phi^6(r)+\frac{m'(r)}{r^2}.
\end{equation}
In case of $\phi=0$, Eq.~\ref{rho-1} reduces to $\tilde{\rho}(r)=\frac{\tilde{m}'(r)}{r^2}=\frac{q^2}{2r^4}-\frac{0.3q^4}{r^8}$ for the EEH black hole in Eq.(\ref{mass-eq}).
Its behaviors are depicted in Fig.~7 (Left) with $q=0.15,0.35,0.5,1,2$, along with the energy density for RN black holes ($\mu=0$).
We observes a negative region near the horizon for $q>0.5$, indicating a violation of the weak energy condition (WEC) for EEH black holes.
However, this violation alone does not imply instability, since violation of the WEC is only one possible signal of instability. 
This is analogous to the negativity of $v(r,0.5,q)$ near the horizon for $q > 0.5$ in EEH black holes (see Fig.~2 (Left)).
On the other hand, for sEEH black holes, negative energy density regions appear near the horizon when $q \leq 0.5$, while for $q > 0.5$ such negative regions shift to an intermediate radial range. 
In all cases, the energy density for sEEH black holes includes negative regions, thus violating the WEC. 
These differences from the EEH black holes arise from the presence of the negative scalar potential.

\section{Dynamical stability for the single branch}

We need to explore a wide range of numerical solutions as functions of $q$ for the single branch in order to perform the stability analysis of sEEH black holes.
Such an analysis is crucial, as it may determine their viability in describing realistic astrophysical configurations with scalar hair.
Our conclusions regarding the stability of sEEH black holes against radial perturbations will be drawn from the examination of the qualitative behavior of the $s$-mode scalar potential as well as the calculation of the QNM frequencies for $s$-mode scalar perturbation.

There are several indications suggesting possible instability in sEEH black holes.
First, the entropy inequality $a_H(0.5,q) < \tilde{a}_H(0.5,q)$ draws our attention, implying that EEH black holes are thermodynamically more favorable than their scalarized counterparts. 
Second, the energy density for sEEH black holes exhibits negative regions for $q > 0$, signaling a violation of the WEC.

Here, we prefer to introduce the radial perturbations around the sEEH black holes as
\begin{eqnarray}
&&ds_{\rm rp}^2=-N(r)e^{-2\delta(r)}(1+\epsilon H_0)dt^2+\frac{dr^2}{N(r)(1+\epsilon H_1)}
+r^2(d\theta^2+\sin^2\theta d\hat{\varphi}^2),\nonumber\\
&&\phi(t,r)=\phi(r)+\epsilon\delta\tilde{\phi}(t,r), \label{p-metric}
\end{eqnarray}
where $N(r)$, $\delta(r)$ and $\phi(r)$ denote the sEEH black hole background, while $H_0(t,r)$, $H_1(t,r)$ and $\delta\tilde{\phi}(t,r)$ represent the perturbations propagating on the numerical background. 
No perturbations are introduced for the gauge field $A_{\hat{\varphi}}$.
From now on, we focus on the $l=0$($s$-mode) scalar propagation, neglecting higher angular momentum modes $(l\neq0)$. 
In this case, the two perturbed metric fields $H_0(t,r),H_1(t,r)$ become redundant via a decoupling procedure, leaving a single linearized scalar equation.

\begin{figure*}[t!]
\centering
\includegraphics[width=0.4\textwidth]{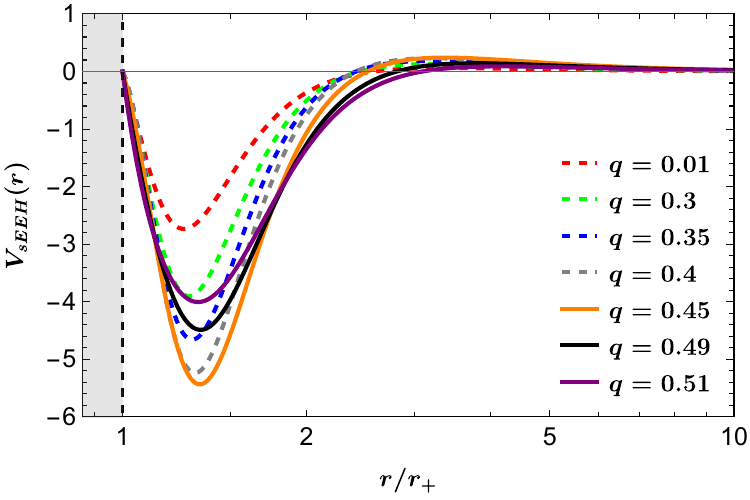}
   \hfill%
\includegraphics[width=0.4\textwidth]{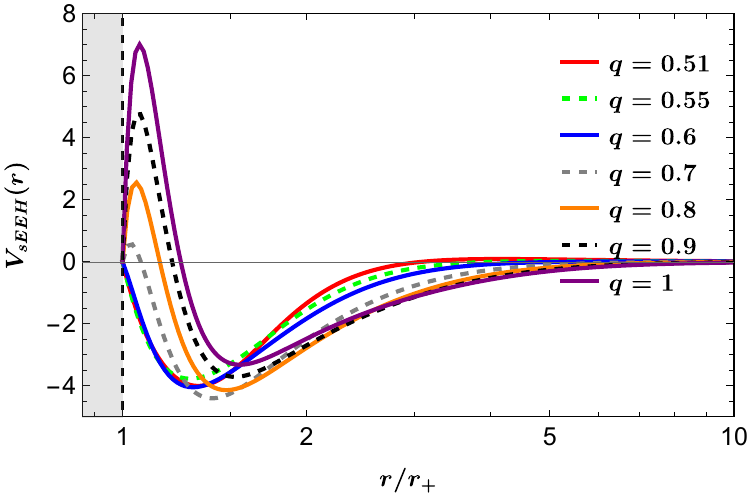}
\caption{(Left) Scalar potentials $V_{\rm sEEH}(r,q)$ for $l = 0$ scalar mode with various values of $q\le0.51$ in the single branch. 
The potential well deepens as $q$ increases and becomes shallower for $q>0.45$.
(Right) As $q$ increases from $q=0.51$, a potential barrier first emerges, followed by the formation of a potential well in the intermediate region.}
\end{figure*}

Considering the separation of variables
\begin{eqnarray}
\delta\tilde{\phi}(t,r)=\frac{\tilde{\varphi}(r)e^{-i\omega t}}{r},
\end{eqnarray}
the $s$-model scalar perturbation reduces to a Schr\"odinger-type equation
\begin{eqnarray}
\frac{d^2\tilde{\varphi}(r)}{dr_*^2}+\Big[\omega^2-V_{\rm sEEH}(r,q)\Big]\tilde{\varphi}(r)=0
\end{eqnarray}
where $r_*$ is the tortoise coordinate defined by
\begin{eqnarray}
\frac{dr_*}{dr}=\frac{e^{\delta(r)}}{N(r)}.
\end{eqnarray}
Here, its scalar potential is given by
\begin{eqnarray} \label{sc-poten}
V_{\rm sEEH}(r,q)=\frac{Ne^{-2\delta}}{r}\Big[N'-N\delta'-2r^2\phi'^2(N'+N/r-N\delta')
-24r^2\phi'\phi^5-30r \phi^4\Big].
\end{eqnarray}
As a consistency check, we verify that setting $\delta(r)=0$, $\phi(r)=0$, and $N(r)\to f(r)$ reduces $V_{\rm sEEH}(r,q)$ to the EEH potential $V_{\rm EEH}(r,M,q)$ in Eq.~\eqref{EEH-P}.

At this stage, we examine the behavior of the effective potential $V_{\rm sEEH}(r,q)$.
Fig.~8 displays scalar potential $V_{\rm sEEH}(r,q)$ for $l=0(s$-mode) scalar with $q>0$ in the single branch.
In the left panel, one observes that $V_{\rm sEEH}(r,q)$ develops negative regions near the event horizon for $q \leq 0.51$.
In contrast, the right panel shows that for $q>0.5$ a potential barrier forms, followed by the emergence of a potential well in the intermediate radial region.
However, the mere presence of a potential well does not, by itself, imply that the single branch with $q>0$ is unstable under radial perturbations.

\begin{figure*}[t!]
   \centering
  \includegraphics[width=0.5\textwidth]{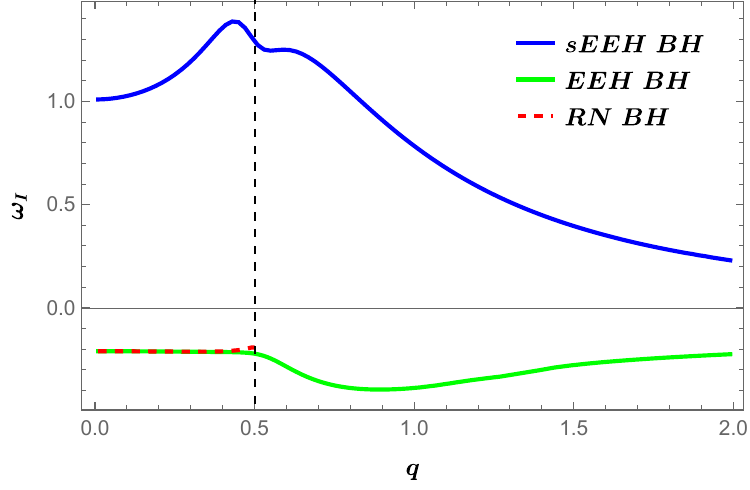}
\caption{Fundamental QNM frequency $\omega_I$ of the $l = 0$ scalar mode as a function of $q$ for sEEH black hole (blue), EEH black hole with $\alpha=0$ (green) and RN black hole with $\alpha=\mu=0$ (red). The black vertical dashed line represents $q=1/2$.}
\end{figure*}

To reach a definitive result, we need to compute QNM frequency defined by $\omega=\omega_R+i\omega_I$ subject to the standard boundary conditions: purely ingoing waves at the outer horizon and purely outgoing waves at spatial infinity.
We determine $\omega$ using the pseudo-spectral method~\cite{Jansen:2017oag}.
Since the real part $\omega_R$ vanishes for all cases considered, only the imaginary part $\omega_I$ is plotted in Fig.~9.
From the numerical results, different with the case of EEH black holes and RN black holes, for the single branch of sEEH black holes with $q>0$, all imaginary parts satisfy $\omega_I(q) > 0$, which signals instability.
Thus, the instability is present for any nonzero magnetic charge $q>0$.
Moreover, $\omega_I(q)$ increases monotonically with $q$ up to a maximum near $q \approx 0.45$, exhibits a transition around $q \approx 0.5$, and then decreases monotonically for larger $q$.
This trend closely follows the variation of the potential well depth in Fig.~8 and is also qualitatively similar to the behavior of the scalar constant $\phi_0(M=0.5,q)$ at the outer horizon, as shown in Fig.~6 (right). On the other hand, we find  from Fig.~9 that RN ($0<q\le 1/2$) and EEH ($0<q\le2$) black holes are stable against radial perturbations because of their $\omega_I(q)<0$, as predicted by (Right) Fig.~2. 

\section{Conclusion and Further Discussions}

We have introduced the EEHS theory to investigate the scalarization of the EEH black hole described by mass $M$, EH parameter $\mu$, and magnetic charge $q$. 
The EEH theory incorporates the one-loop effective Lagrangian of QED coupled to the Einstein gravity~\cite{Yajima:2000kw}.
It is worth noting that for $\mu > 0.019$ at $M=0.5$, there is no restriction on the magnetic charge $q$ for the existence of a single horizon.
In the limiting case $\mu=0$, the solution reduces to the RN black hole, which possesses both outer and inner horizons.
For the present study, we set $\mu=0.3$ so that the spacetime admits a single horizon.

To realize the simplest scalarization of the EEH black hole, we introduce a negative single term of the scalar potential $V(\phi)=-4\alpha^2 \phi^6$~\cite{Chew:2024evh}.
The onset scalarization does not occur for this black hole because it remians stable under scalar perturbations.
However, the introduction of a negative potential can trigger scalarization, thereby providing a novel mechanism for evading the no-hair theorem\cite{Herdeiro:2015waa}.
By numerically solving the full set of three field equations, we obtained a single-branch family ($q>0$) of sEEH black holes.
Notably, there is no existence line separating scalarized and non-scalarized configurations in this case. A significant characteristic of sEEH black holes is that the scalar hair exhibits primary charge behavior for $q<1/2$ and secondary charge behavior for $q>1/2$ at a fixed ADM mass ($M = 0.5$). This distinctive property highlights a novel feature of these black holes.

Several indications point toward the instability of the sEEH branch.
First, we note the entropy inequality $a_H(0.5,q)<\tilde{a}_H(0.5,q)$ which suggests that the EEH black hole is thermodynamically preferred over its scalarized counterpart.
Second, the energy density of sEEH black holes exhibits negative regions for $q>0$, signaling a violation of the WEC.
We have further examined the stability of sEEH black holes by analyzing the $s$-mode scalar potential and computing the QNM frequencies under radial perturbations.
In all cases with $q>0$, the imaginary part of the QNM frequency satisfies $\omega_I(q)>0$ indicating that the single branch of sEEH black holes is dynamically unstable against linearized scalar perturbations.

It is worth noting that our results, as the simplest attempt at scalarization in the EEH gravity theory, highlight its substantial potential in the study of black hole physics.
The scalar charge exhibits a slight variation from a constant value with increasing $q$, but undergoes a rapid increase as the black hole enters the overcharging regime, displaying a distinct property from primary to secondary hair characteristics. 
This behavior highlights the intricate and nontrivial structure inherent to the EEH theory. 
It further suggests that black holes in the overcharging regime, where the charge-to-mass ratio exceeds the extremal bound, may exhibit phenomena intimately connected to the principles of quantum gravity~\cite{Javed:2024hzs,Sorokin:2021tge} and the cosmic censorship conjecture~\cite{Hubeny:1998ga,Liu:2020cji,Alipour:2025dan,Kehagias:2023qmy,Shaymatov:2023gfh}.
Such connections are particularly significant for advancing our understanding of possible extensions or evasions of the classical black hole no-hair theorem.
In this sense, the EEH theory offers a fertile testing ground for exploring the interplay between strong-field gravity, nonlinear electrodynamics, and quantum effects in extreme spacetimes.

\vspace{1cm}

{\bf Acknowledgments}

 Y.S.M. was supported by the National Research Foundation of Korea (NRF) grant funded by the Korea government(MSIT) (RS-2022-NR069013). M.P. and H.G. were supported by the Institute for Basic Science (Grant No. IBS-R018-Y1). We appreciate APCTP for its hospitality during completion of this work.
 \vspace{1cm}


\end{document}